\documentclass[preprint,12pt]{elsarticle}
\usepackage{graphicx}
\usepackage{amssymb}
\usepackage{lineno}

\usepackage{float}

\journal{Astronomy and Computing}

\begin{document}

\begin{frontmatter}


\title{Cloud access to interoperable IVOA-compliant VOSpace storage}

\author[oat]{S.~Bertocco}
\author[canfar]{P.~Dowler}
\author[canfar]{S.~Gaudet}
\author[canfar]{B.~Major}
\author[oat]{F.~Pasian}
\author[oat]{G.~Taffoni}
\address[oat]{INAF-OATs, Istituto Nazionale di Astrofisica - Osservatorio Astronomico di Trieste, Trieste, Italy}
\address[canfar]{CADC, National Research Council Canada, Victoria, BC, Canada}

\begin{abstract}
Handling, processing and archiving the huge amount of data produced by the new generation of experiments and instruments in Astronomy and Astrophysics are among the more exciting challenges to address in designing the future data management infrastructures and computing services.We investigated the feasibility of a data management and computation infrastructure, available world-wide, with the aim of merging the FAIR\footnote{Findable, Accessible, Interoperable, Reusable} data management provided by IVOA standards with the efficiency and reliability of a cloud approach. Our work involved the Canadian Advanced Network for Astronomy Research (CANFAR) infrastructure and the European EGI federated cloud (EFC). 
We designed and deployed a pilot data management and computation infrastructure that provides IVOA-compliant VOSpace storage resources and wide access to interoperable federated clouds. In this paper, we detail the main user requirements covered, the technical choices and the implemented solutions and we describe the resulting Hybrid cloud Worldwide infrastructure, its benefits and limitations.
\end{abstract}

\begin{keyword}
IVOA \sep Virtual Observatory \sep Cloud \sep Interoperability \sep Distributed services  \sep VOSpace  \sep Storage
\end{keyword}

\end{frontmatter}


\section{Introduction}
\label{intro}
Astronomy and astrophysics (A\&A) has become a BigData science thanks to the advent of numerous experiments across the electromagnetic spectrum 
with many terabytes of images and with billions of detected sources, and often with tens of measurements for each of the observed objects (e.g.  \citep{planck} \citep{wmap} \cite{SloanSurvey} \cite{Sabater2017} \cite{2007ASKAP}  \cite{magic}).
Moreover, during the next decade, new highly complex and massively large data sets are expected to be provided by a new generation of scientific instruments implemented by large international collaborations (e.g. the Square Kilometer Array~\cite{ska}, the Cherenkov Telescope Array~\cite{cta}, the Extremely Large Telescope E-ELT~\cite{eelt},  the James Webb Space telescope~\cite{jwst},  the Large Synoptic Survey Telescope \cite{lsst}, or the Euclid satellite mission~\cite{euclid}). 
Finally, high-resolution numerical simulation codes are producing in-silico experiments resulting in PetaBytes of data to be stored and analyzed. 
 
Thus, the huge data production foreseen by this new generation of experiments and their technical complexity requires the adoption of new approaches that overcome the traditional research methods,  of new computing and storage facilities, of new science platforms, and algorithms for the access, visualization and analysis of astronomical data.

In the last decades  scientific communities adopted production e-Infrastructures or Distributed Computing Infrastructures (DCI) to satisfy their computing and storage requirements (e.g. the European Grid Infrastructure EGI\footnote{http://www.egi.eu}
and the Open Science Grid\footnote{http://www.opensciencegrid.org}) and to manage large distributed collaborations.

In the framework of DCIs, Cloud Computing (CC) has emerged as one of the dominant technologies, because it offers 
flexibility in computing and data resources exploitation.
 
The use of Cloud Computing facilities is  thus becoming increasingly important 
also within the  A\&A community. For example in ~\citep{Sabater2017}, \citep{cloud2} or \citep{Berriman}, the authors showed some   interesting results in terms of simplicity for the software installation and maintenance, flexibility to adapt the infrastructure to the needs of the problems, scalability of the infrastructure due to on-demand consumption of (shared) resources and services accessibility (standard APIs, services and tools).

While CC technologies allow to create open platform environments oriented to data and computing for experiments, the A\&A community needs not only traditional computing resources but also the use of complex data operations that require on-line access to catalogs and archives.

Traditionally A\&A has been at the forefront of implementing digital repositories for preserving and curating their own data. Typically such repositories are maintained within data centers with appropriate provision for tools/services for access and analysis. Data archives are expanding rapidly, e.g. through flagship high data volume generating A\&A projects, and EuroVO\footnote{Census of European DataCentres http://goo.gl/zOWykc} has recently identified over 70  data centres in Europe.

It is  also well-known that since more than forty years FITS\footnote{https://fits.gsfc.nasa.gov/fits\_home.html}
has been used as a standard for data exchange, and since the 2000s, the Virtual Observatory (VObs)
has been THE standard for  A\&A data interoperability. The VObs is a collection of distributed interoperating data archives and software tools using the Internet in order to form a scientific research environment in which  astronomers are able to conduct astronomical research programs finding, accessing and processing data seamlessly, regardless of their physical location.
The VObs distributed data oriented infrastructure is based on standards, tools, software and services defined by the International Virtual Observatory Alliance (IVOA)\footnote{http://www.ivoa.net}, a world-wide organization with the task of facilitating the international coordination and collaboration to release standards and tools.
A\&A data centers are employing VObs to provide seamless unified access to distributed and highly heterogeneous data archives.

The VObs is  a data and services platform that complements the distributed computing and storage infrastructures (including CC).  
Integrating the two technologies into a single framework would allow to  build a science platform for A\&A users and projects. This approach  as been already recognised as an important technology solution, indeed 
EuroVO emphasises  that  the relation between the VObs and Cloud technologies should be taken into appropriate consideration in order to provide 
a complete data usage ecosystem for A\&A communities\footnote{EuroVO--CoSADIE Project http://www.eurovo.org}.

To this extent, A\&A community has gathered rich experiences  within the CANFAR (Canadian Advanced Network For Astronomy Research) federated cloud deployed on Compute Canada resources and operated by National Research Council Canada. CANFAR is  a unique example of an A\&A oriented infrastructure that joins together the Infrastructure as a Service (IaaS) cloud and the standards and services developed by the IVOA.
However, to offer a science platform for the new generation of experiments it is necessary to design an infrastructure  involving geographically distributed,
international  Cloud facilities. 

In this paper we present our work  to provide a new innovative science platform built for European astronomers and Astronomical Data Centers,
integrated in the EGI Federated Cloud, providing A\&A users with wide access to data, storage and computation resources.
The aim of this work is  building an Hybrid Cloud that interconnects the CANFAR cloud infrastructure and the 
EGI Federated Cloud,  both remaining unique entities, owned, managed  and operated by different institutions
but bounded together by means of  a common authentication model and by the IVOA standards implementation.
To reach this goal we:
\begin{itemize}
\item exploit  IVOA standards to ensure FAIRness (Findability, Accessibility, Interoperability, Reusability) of data and at the same time of the computing resources management;
\item exploit and extend a set of software APIs, open source released by the Canadian Astronomy Data Center (CADC), to set up a set of services interoperable with the CANFAR one, in Europe;
\item provide geographically distributed storage services, ensuring a wide networked accessibility of data and flexibility in resource usage, moving a new step toward the realisation of the VObs and of an international distributed A\&A facility;
\item provide a uniform and seamless access to heterogeneous resources necessary to organize and process high data volumes.
\end{itemize}

This work is exploring a possible technological solution for large projects,  to implement a new integrated approach to data access, manipulation and sharing. In particular,  in case of world wide distributed collaborations that needs to share distributed infrastructures.

The paper is organized as follows. In section 2  we introduce the Virtual Observatory and IVOA; in section 3 we describe Cloud Computing in general and we detail the CANFAR and EFC platforms implementation. Section 4 presents the developments of our science platform in the context of the Virtual Observatory framework to provide an integrated infrastructure for big data analysis, management and visualization; we detail the technical implementation and the different aspects to face (authentication, authorization implementation policies, interoperability aspects).
Finally we present actual applications and future perspectives of our work. 

\section{The Virtual Observatory and IVOA Standards}
\label{IVOA}

Astronomy produces largely increasing amount of data of many kinds, coming from various sources: science space missions, ground based telescopes, theoretical models, compilation of results, etc. Traditionally data are managed both by large data centers and smaller teams providing the scientific community with data and/or computing services through the Internet. This is a Resource Layer. Individual researchers, research teams or computer systems need to access these data interacting with a User Layer, i.e. a set of access, management and analysis facilities. The Virtual Observatory (VObs) is the necessary “middle layer” framework connecting the Resource Layer to the User Layer in a seamless and transparent manner~\citep{2010ivoaRept1123A}.
Like the web, which enables end users and machines to access transparently documents and services without concern about where they are stored, the VObs enables the astronomical scientific community to access astronomical resources without bothering where they are stored by the astronomical data and services providers. The VObs makes available a technical framework for the providers to share their data and services (``Interoperating''), allowing users to find (``Finding'') these resources, to get them (``Accessing'') and to use them (``Re-Using''). In this sense VObs implements a FAIR data management.
The International Virtual Observatory Alliance (IVOA) is an organisation that debates and agrees the technical standards that are needed to make the VObs possible. 

In the A\&A research community the standards and tools delivered by the IVOA are widely used for data management (intended as data storage, archival and preservation) and for data analysis. 
To provide A\&A researchers with an engaging and useful operative environment, cloud infrastructures should make available access to storage, data and computing in a VObs compliant manner, i.e. through IVOA tools and services. 

\section{Cloud Computing e-infrastructures}
\label{cloud}
Cloud Computing in its simplest sense, is a computing and storage resources management model. 
It is a method for pooling and sharing hardware infrastructure resources on a massive scale: finite hardware resources are clustered using a Cloud Management Stack and shared, between competing demands, through the on-demand flexible provision of services or virtualized computing, networking and data storage capabilities. 
In a Cloud, each user has the illusion of exclusive control over the underlying environment without needing to know anything about how the physical resources are configured.

Cloud resources delivery can be divided into three services layers, as proposed
by the US National Institute of Standards and Technology\footnote{https://csrc.nist.gov/publications/detail/sp/800-145/final}~\citep{mell2011nist}, namely:
Software as a Service (SaaS) , Platform as a Service (PaaS) and Infrastructure as a Service (IaaS).
At the highest level, SaaS allows users to use the provider’s 
applications running on a cloud infrastructure (e.g. Google suite).
In a PaaS cloud the user can deploy his own software applications in the 
cloud infrastructure, while at the lowest level, an IaaS provides users with virtualized storage, processing, and other computing resources.

Cloud technology is also characterized by four deployment models:
\begin{itemize}
\item private cloud, exclusively used by an organization; 
\item community cloud, used by a specific community of consumers from organizations that have shared concerns (e.g., mission, security requirements, policy, and compliance considerations); 
\item public cloud, for general public open use; 
\item hybrid cloud, a composition of two or more distinct cloud infrastructures (private, community, or public) tied together by standardized or proprietary technology which enables data and application portability. 
\end{itemize}

In this paper we describe our design and implementation of an IaaS Hybrid Cloud based on CANFAR and EFC (EGI Federated Cloud).

\subsection{EGI Federated Cloud}
\label{EGIFedCloud}
The EGI Federated Cloud\footnote{https://www.egi.eu/federation/egi-federated-cloud/} is an European open cloud system that offers a scalable and flexible e-infrastructure to the European research communities. The Federation pools IaaS, PaaS and SaaS services from a heterogeneous set of cloud providers using a single authentication and authorization framework that allows the portability of workloads across multiple providers.

\begin{figure}[ht]
\centering\includegraphics[width=1.0\linewidth]{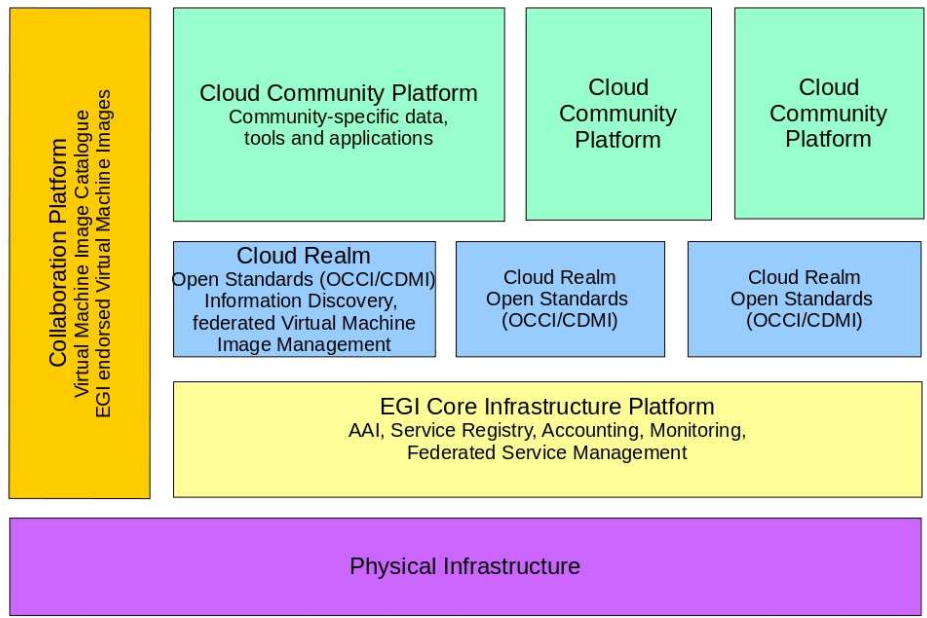}
\caption{EGI Federate Cloud Layered Infrastructure.}\label{fig:EgiArchFig}
\end{figure}

The EFC architecture defines cloud specific capabilities and interfaces and a number of core services for operations and security. Those services can be used to build community platforms and cloud Realms (See Figure~\ref{fig:EgiArchFig}).
Every participating cloud provider has got to expose to users and system administrators the same set of interfaces(\citep{OCCI}, \citep{CDMI}), 
so enforcing cloud technology agnosticism. 
This way EGI supports more than one Cloud stack implementation exposing a set of common interfaces. 
In our work we use OpenStack Mitaka, that is the same platform used by CANFAR, however EFC supports also OpenNebula and Synnefo.

The core services required to build the Federated Cloud, are: Federated Authentication and Authorization Interface (AAI) which allow users to connect to all cloud sites using the same credentials, the Accounting which collects, integrates and displays usage information across the whole federation, the Information System which discovers the available resources in the infrastructure, the Monitoring service which monitors services and resources status and availability, the Service Registry for configuration management of federated cloud services and Service “marketplace” where free and paid resources can be listed and discovered. Finally, the EGI Applications Database (AppDB) is a central service that stores and provides to the public information about software solutions in the form
of native software products and/or virtual appliances. Some registered solutions enable users to deploy and manage Virtual Machines to the EGI Cloud infrastructure automatically.

Within EGI, research communities are generally identified and, for the purpose of using EGI resources, managed through ``Virtual Organisations''. The users are identified with X.509 certificates which are extended with Virtual Organisations attributes as discussed in Section~\ref{AA_and_interop}.

\subsection{CANFAR Infrastructure}
\label{CANFAR}
\begin{figure}[ht]
\centering\includegraphics[width=1.0\linewidth]{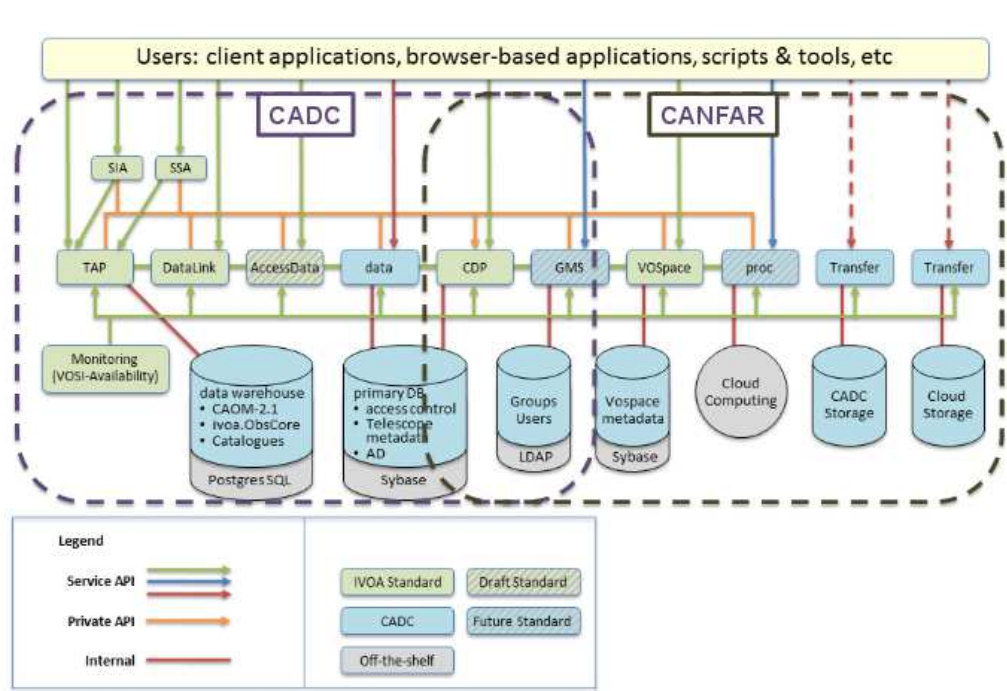}
\caption{CANFAR infrastructure schema. The main components of the infrastructure and the connections between the IVOA services offered by CADC and the CANFAR computing service are sketched.}
\label{fig:CanfarArchFig}
\end{figure}

CANFAR is an integrated e-infrastructure to support collaborative astronomy (A\&A) teams, especially those with data intensive projects. 
It combines the Canadian national research network (CANARIE), the cloud processing and storage resources of Compute Canada and an astronomy data center, the Canadian Astronomy Data Center (CADC), into a unified service. 
CANFAR is developed, maintained and operated by the CADC with contributions from several Canadian universities. In Figure~\ref{fig:CanfarArchFig}, we sketch the CANFAR 
architecture and the connections between the IVOA services and the computing services. CADC is offering both private and public data access to astronomical data archives using IVOA standards. An authentication and authorization service is used to provide access to proprietary data. The same service is used by CANFAR to allow the access to computing and storage resources and to services for proprietary data analysis.

CANFAR has four principal user-facing capabilities, namely: user storage, virtual machines on demand, batch processing and groups management. All are RESTful web services accessible via http or via web user interfaces. 

The authentication is based upon username/password and a X.509 certificate mechanism and the services interoperability is granted by the IVOA credential delegation protocol \citep{2010ivoa.spec.0218P} implementation. Using this protocol, X.509 user's credentials can be delegated so, using them, a service can make requests of other services in the name of that user.

The authorization is based on a group membership policy. A RESTful web service to manage groups and memberships, called Group Management Service (GMS), is available that guarantees a fine grained authorization system. 
It supports the following operations: creating, getting, updating and deleting a group; adding and deleting a user to/from a group; adding and deleting a group to/from a group.

A distributed storage layer allows users to access and share their data. It is based on a VOSpace IVOA service \citep{2013ivoaSspec0329G} and on a data transfer service. CANFAR provides also a FUSE  VOSpace implementation to allow users to mount their VOSpace on virtual machines. 

CANFAR is offering a cloud computing facility that allows the creation of on demand virtual machines and data-location-aware virtual clusters to process and analyze data. 
The system is based on standard OpenStack technologies (e.g. Nova) and IVOA standards that allow the execution and management of jobs in the cloud (the IVOA Universal Worker Service \citep{2016ivoa.spec.1024H}).

Each web service supports a RESTful monitoring interface that complies with the VOSI-availability  standard \citep{2017ivoa.spec.0524G}. These interfaces can be monitored via periodic requests from many standard monitoring packages to provide a central view of the system. 

\section{Building an Hybrid Cloud: the architectural view}
\label{description}

In our work we designed and deployed an hybrid cloud gathering CANFAR and EFC  cloud infrastructures and we identified services and interfaces required to make the two clouds cooperate and interoperate. 
Our aims was to allow A\&A data and applications to work in the same way regardless the infrastructure and to implement the ability of data, VMs and software to be easily moved and reused in the two cloud environments. 
\begin{figure}[ht]
\centering\includegraphics[width=.5\linewidth]{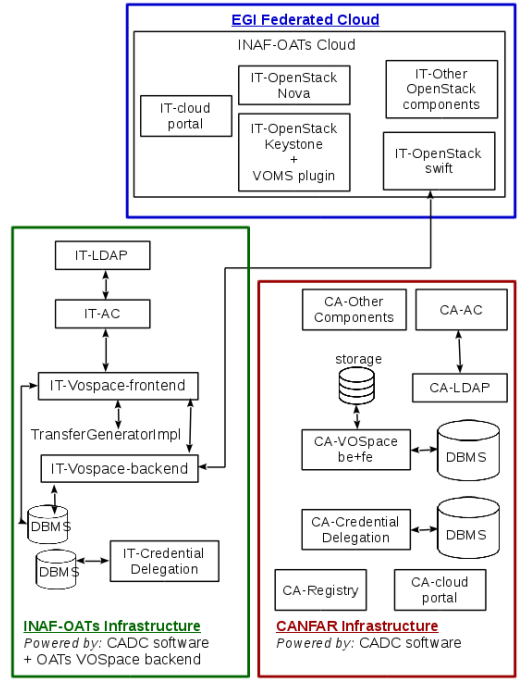}
\caption{General architecture. The three main blocks of the science platform are depicted: CANFAR, EGI and the gateway site INAF-OATs. CANFAR and the gateway are interoperable thanks to the IVOA standards implementation. The gateway is EGI compliant short-circuiting the interoperability. The more significant software components are highlighted.}\label{fig:archFig}
\end{figure}
There are only a few essential building blocks towards an interoperability  for IaaS \citep{interop1, interop2} and we focus 
our attention on three of them in particular: access mechanisms, security and storage. 

By implementing the Authentication and Authorisation Infrastructure (AAI) federation, users  exploit transparently resources provided via the two clouds.
An interoperable  AAI allows European users to access CANFAR resources 
(data and computing) using their European (EGI Federated Cloud) credentials and vice versa. 
An interoperable storage allows data sharing (e.g. to replicate open and private data for data availability and preservation) and Virtual Machines (VMs) sharing.
Astronomers or developers can access Cloud services for VMs, Virtual appliances or application executions using APIs or Command line interfaces 
offered by OpenStack which is a common "middleware" for  both  EFC and CANFAR.

The federation model proposed in this document is based on the assumption that the two clouds will remain independent and independently managed but  bounded together by a set of interoperable services based on IVOA standards.
The interoperability is achieved through an OpenStack gateway site, hosted by the INAF Trieste Astronomical Observatory (INAF-OATs), that implements the interoperability services.
This site is federated in the EFC, which is offering a multi-community cloud providing generic services to make available Community Platforms \citep{EGIFedCloud} as shown in Figure~\ref{fig:EgiArchFig}.

The INAF-OATs gateway hosts a deployment of the services and interfaces implementing the A\&A platform. These services are provided as EGI Community Services and based on IVOA standards:
\begin{itemize}
\item the Access Control Service is the AAI service implementation managing users and groups.  
\item the VOSpace service  is a distributed storage implementation\footnote{In this work we adopt the VOSpace 2.1 proposed recommendation as described by http://www.ivoa.net/documents/VOSpace/20170924/index.html};
\item the Credential Delegation Service is the IVOA Credential Delegation Protocol implementation~\citep{2010ivoa.spec.0218P}. 
\end{itemize}
Those services are also available to the EFC users thanks to  the Service Registry and the Market place. 
A VM image, configured to support a pool of ad hoc clients and analysis tools (ESO Scisoft package, TOPCAT), is available in the EGI AppDB.

A  picture of the main building blocks used to build our hybrid cloud and to reach interoperability and compatibility is  presented in  Figure~\ref{fig:archFig},
the role of the gateway site and its services is highlighted. The green box isolates the Community services deployed in the gateway and their role, 
including their persistence layer.

The Access Control (AC) guarantee a seamless access to both EGI and CANFAR infrastructures to users identified by  their X.509 certificate while the services interoperability is achieved exploiting  the IVOA credential delegation protocol~\citep{2010ivoa.spec.0218P}. 
For what regards the authorization,  EFC implements a coarse grained approach based on Virtual Organisations while CANFAR, OATs-INAF gateway site and consequently our Community Cloud  Platform, implements a fine grained approach based on  the data ownership and on the user group memberships.
In EFC, Astronomers will be mapped in a single specific Virtual Organization and the granularity of local data access policies is driven by the local groups memberships stored in the local AC Service at the gateway site.

Each service has a persistence layer.
The Access Control persists data on an LDAP database. We have chosen LDAP because widely used in data and computing centers to store users and groups informations for authentication and authorization purpose. This allows already existing LDAP databases to be integrated with  or directly used by the Access Control service.

The VOSpace Service uses a relational database to save meta-data associated with stored files while we implement a  vospace-backend service and a  TransferGenerator on the gateway site in order to interface  with a EGI specific storage solution (Openstack Swift).
The Credential Delegation Service uses a relational database  to store the user's delegated credentials. 

We implemented different solutions for the persistent layer: Sybase at CANFAR and MysqlDB at EGI. 
We also made a number of  software customisations to make the CADC provided APIs more flexible in supporting different persistent layers.

In the following sections we are going to detail the software components implemented by the CADC open source cloud project or by the INAF-OATs development group and how they have been deployed to achieve the federated environment building our science platform.

\section{Software description and functionalities}
\label{sw_desc}
The software used to deploy the INAF-OATs gateway site is a set of java RestFul web services. It is mainly released by the CADC in an open source repository\footnote{https://github.com/opencadc}. 

In our work, we adapted and deployed three CADC released modules to achieve interoperability: Access Control, Credential Delegation and VOSpace interface. INAF-OATs releases a vospace-backend software module implementing a storage management service\footnote{https://github.com/oats-cadc} which, integrated with the VOSpace interface,  makes available a working storage service.

\paragraph{Access Control}
\label{ac}
The Access Control Service stores the information about users and groups and provides authentication mechanisms using CADC released software APIs. The CADC User Authentication Model, adopted by INAF-OATs also, is a model for representing users and groups. It is implemented in Java using the Java Authentication and Authorization Service (JAAS). 
A user is internally uniquely identified by one numerical identity, but he can have a number of other identities, for different contexts, aligned with the IVOA Single-Sign-On Profile~\cite{2017ivoaSpecSSO}: an HTTP-identity, which  is the web user identity associated with the simple HTTP user-password access; a X.509 identity, which is a X.509 certificate identity; a Cookie identity, which is a web user identity associated with a cookie. The set of supported identities can be increased with new kinds of identities by design.

The user registration follows a classic registration mechanism added with a user registration authorization process: the user provides to the service some registration information, the registration request is put in standby, a service administrator must check the request and decide if accept or reject it. This is the implementation of a common requirement in scientific platforms allowing administrators to check pertinence of the registration request and in case, to configure authorization policies for the new user contextually with the registration authorization.
Groups represent associations of users. They have an owner, administrative members, and actual members. 

\begin{figure}[ht]
\centering\includegraphics[width=.9\linewidth]{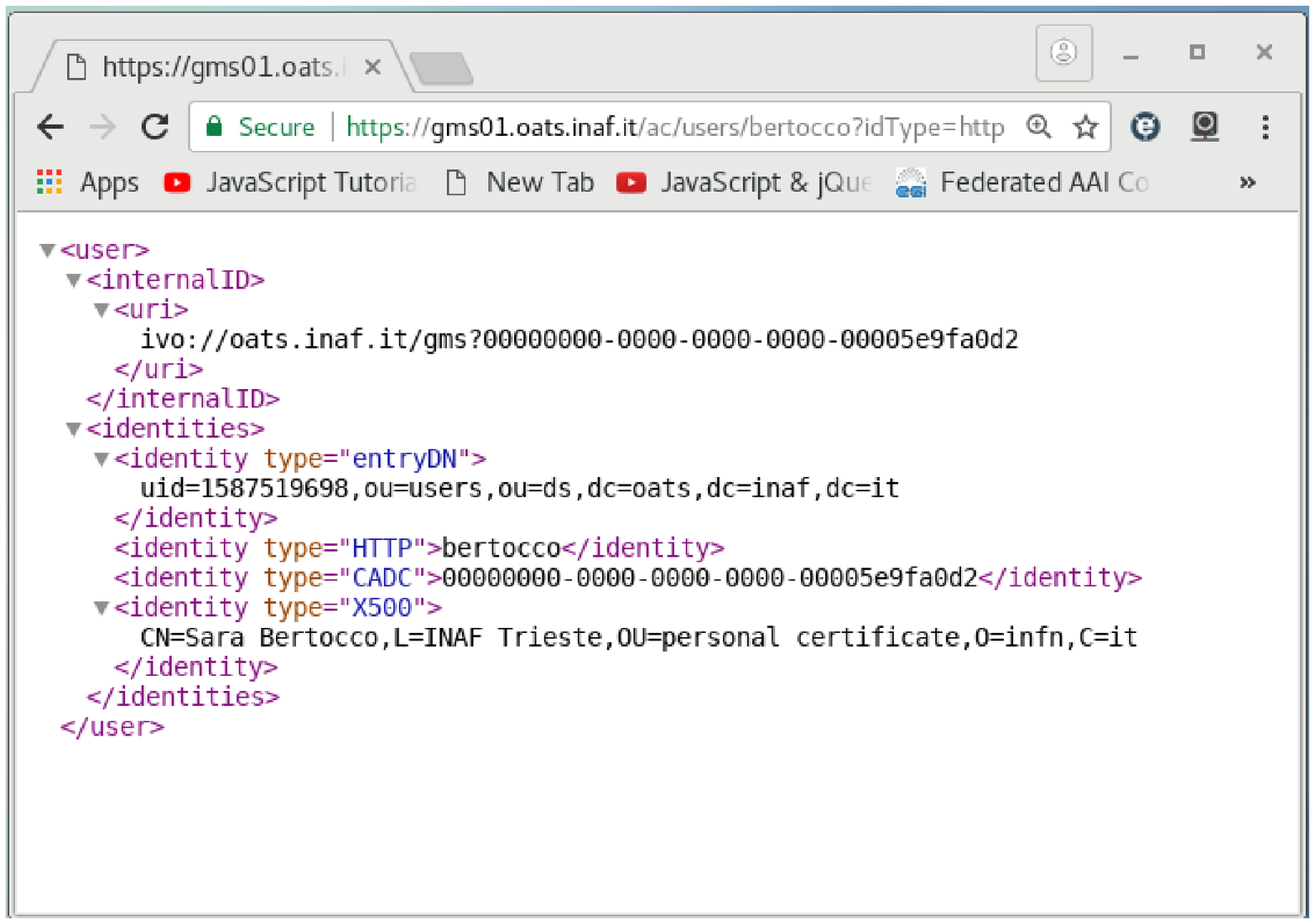}
\caption{XML response to a user’s information request.}\label{fig:user_info}
\end{figure}

\begin{figure}[ht]
\centering\includegraphics[width=.9\linewidth]{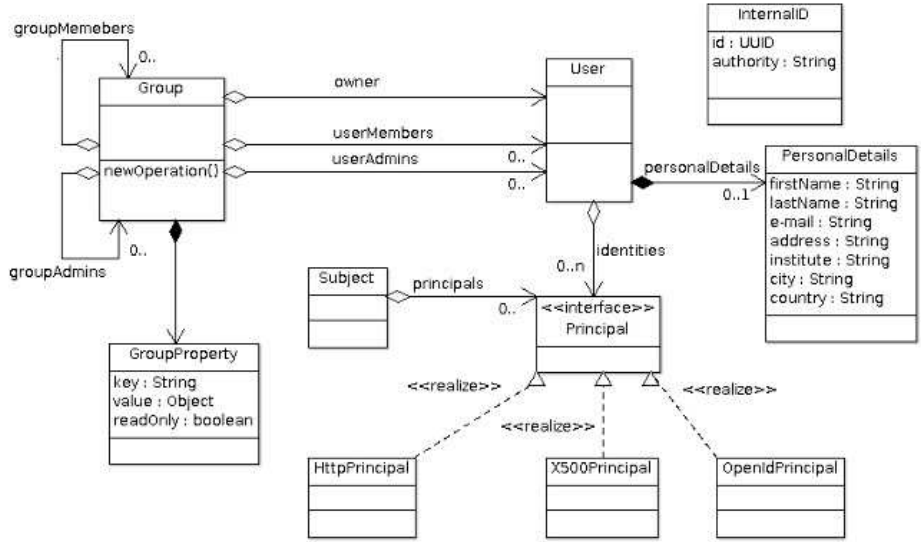}
\caption{CADC User Authentication and Authorization Model Class Diagram.}\label{fig:ac_class}
\end{figure}

The management of users and groups is performed by  RESTful web services using an LDAP server as persistence layer. This service can be queried to ask informations about user's identities and groups memberships then used as authorization informations (see figure~\ref{fig:user_info}). 
The class diagram of the user authorization model adopted is in Figure~\ref{fig:ac_class} and the authorization mechanism is described in section~\ref{AA_and_interop}.
\paragraph{Credential Delegation Service}
\label{cdp}
Credential Delegation Service is an IVOA  RESTful web service that enables users to create temporary credentials at a site enabling that site to perform web service actions on their behalf. This feature is the key to allow the interaction among services located at different sites. 

The software implementation is based on AstroGrid security-facade library~\cite{astrogrid}~\cite{astrogridSecurity}, which provides a Java implementation of both the service and client ends of the IVOA Credential Delegation Protocol~\citep{2010ivoa.spec.0218P}. Certificate and proxy validation is performed using Bouncy Castle\footnote{Bouncy Castle is a collection of APIs used in cryptography. It includes APIs for both the Java and the C\# programming languages}. The service end point of this implementation is a servlet which, by default, caches the delegated credentials in memory. It's designed to be plugged into any Java web-application and extended to use any persistence layer as static credential store. The CADC Credential Delegation Service includes the implementation of a persistence layer storing credentials in a relational database. We modified the implementation to make the software flexible enough to allow the use of different databases (we used sybase, mysql, mariadb and postgress  in different environments: CADC, INAF-OATs and testing in both sites). 

\begin{figure}[ht]
\centering\includegraphics[width=\linewidth]{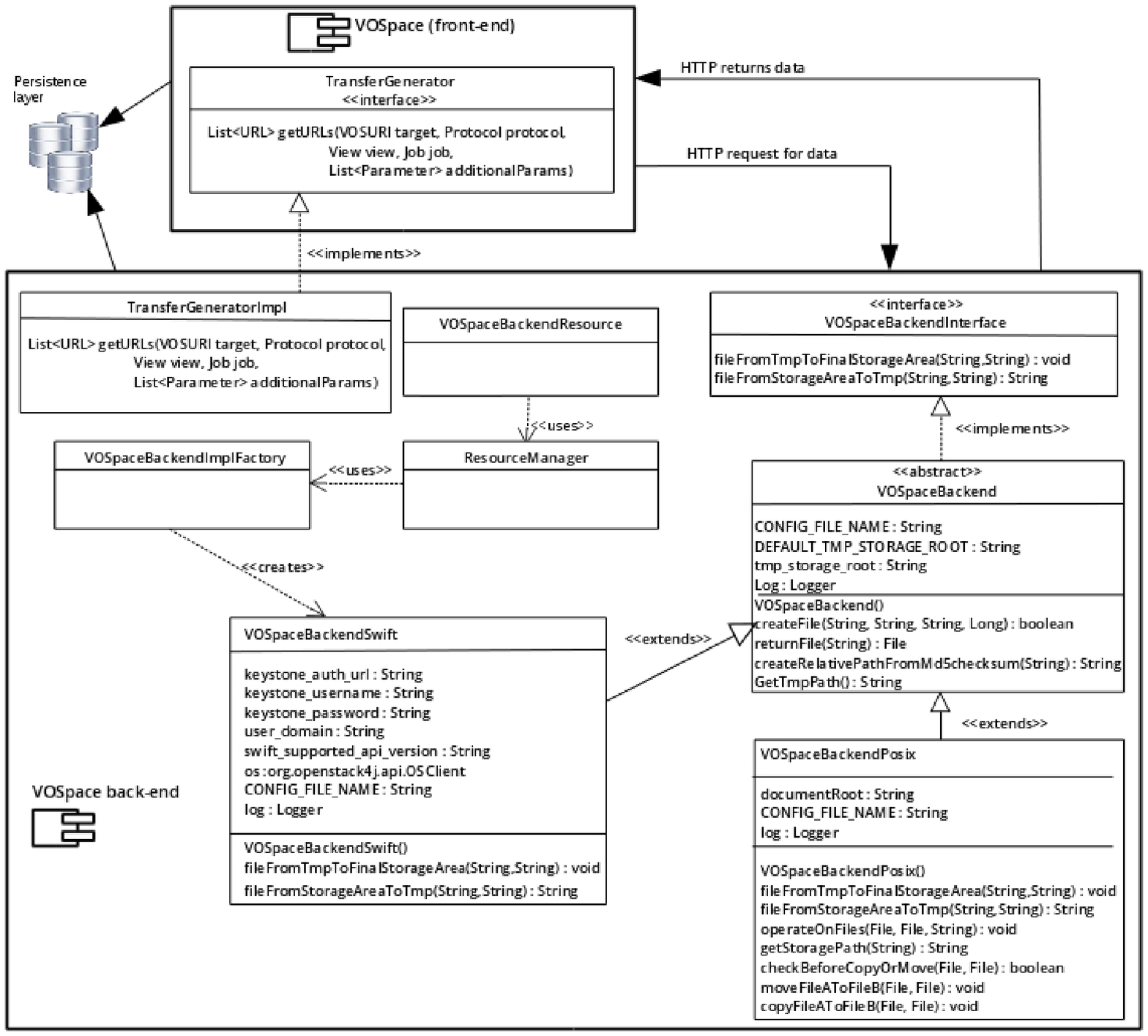}
\caption{VOSpace Services and Class Diagram.}\label{fig:vos_class}
\end{figure}

\begin{figure}
\centering\includegraphics[width=.9\linewidth]{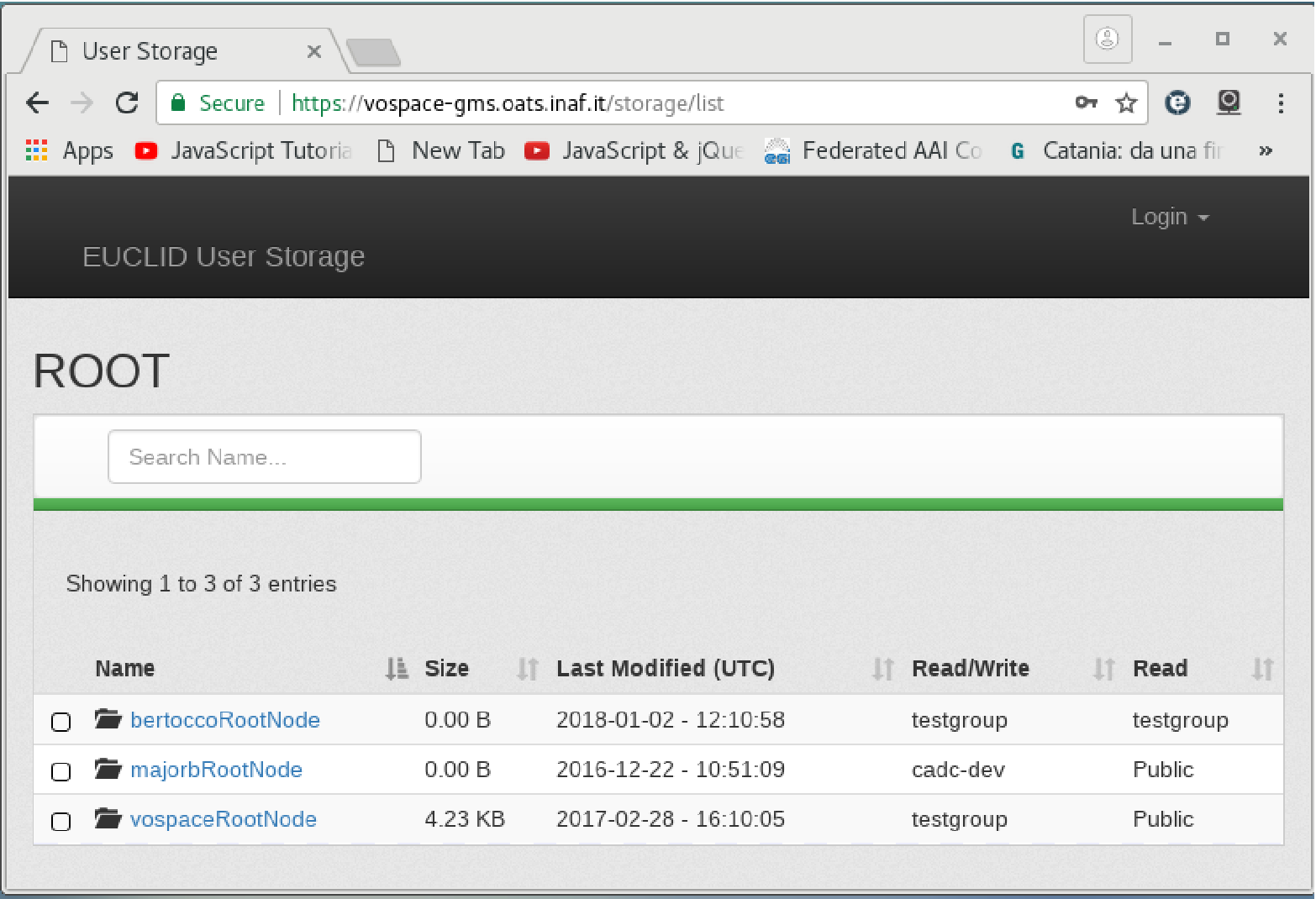}
\caption{VOSpace web client main interface.}\label{fig:web_client}
\end{figure}

\paragraph{VOSpace}
\label{vospace}
VOSpace exposes to users an interface to storage to allow them to access, process, and share their data. It conforms to the VOSpace IVOA recommendation~\cite{2013ivoaSspec0329G}. 
The IVOA recommendation defines the control messages required to access the data, and not how data is stored or transferred. Thus, the VOSpace interface can readily be added to an existing storage system.
On this base, we splitted the implementation of a storage service in three main components: the VOSpace interface, here called VOSpace front-end, responsible of the meta-data management; the data transfer service, responsible for handling the upload and download of data to and from their physical storage location; the vospace-backend, in charge to manage the physical storage layer and to point  to the data location. Figure~\ref{fig:vos_class} shows the software modules with their main classes, relationships and persistence layer. 
The VOSpace front-end has different implementations: a command-line Java client, a web client shown in figure~\ref{fig:web_client} (Java and Javascript written) and a command-line Python client. 
The vospace-backend links the VOSpace metadata structure, called Node, with the stored data identified in order to keep the information about physical data location. The storage solution can be changed thanks to a plug-in architecture allowing to plug-in at run time different storage solutions. The current implementation provides two plug-ins: one for a posix file system storage and one for an OpenStack swift storage implementation. The selected plug-in has to be shaped in a configuration file.

\section{Authentication, authorization and interoperability}
\label{AA_and_interop}

The services interoperability is granted by the use of X.509 personal certificates for user authentication both in Virtual Observatory services and EGI Federated cloud sites. 

To exploit the interoperability of services, registered users must have a X.509 certificate identity. At CANFAR, an internal Certification Authority (CA) automatically issues a X.509 certificate to users. However, a user already having a personal certificate can delegate that certificate by uploading a proxy to the Access Control Service. At OATs-INAF, the user needs to have a certificate issued by a Certification Authority (CA) member of the Interoperable Global Trust Federation (IGTF). All the Authorities releasing certificates must be trusted on both sides of the platform: EFC and CANFAR.

The X.509 certificates are accepted by the EFC also. 
Within EGI, research communities are generally identified and, for the purpose of using EGI resources, managed through ``Virtual Organisations''. Users are identified with X.509 certificates extended with Virtual Organizations attributes (e.g. acknowledging that a user is member of a Virtual Organization) in a so called VOMS proxy certificate. VOMS (Virtual  Organization  Membership  Service) is a well proven technology, to provide identity federation across different providers~\cite{VomsId}. The EFC OpenStack based endpoints, suitably configured using an integration module for VOMS authentication named keystone-voms\footnote{Keystone-VOMS is an extension module of Keystone, the OpenStack service that provides API client authentication, service discovery, and distributed multi-tenant authorization. Keystone-voms enables OpenStack authentication of users through X.509 certificates with VOMS extensions. It is designed to be integrated as an external authentication plug-in, so that Keystone will preserve its original features and users will still be able to authenticate using any of the Keystone native mechanisms}, are able to map the voms proxy certificate with the Virtual Organization information in users authentication and authorization mechanisms in OpenStack. 

The OATs-INAF cloud site is configured to support VOMS proxy authentication, so the VOMS proxy certificates can be used to query both the Virtual Observatory services and the EFC sites suitably configured with the keystone-voms module. Users of the IVOA services, to be authenticated and authorized, must be registered in a common Virtual Organization which is mapped in a specific Domain in the OpenStack-Keystone service.

To well clarify the authorization mechanism, in the Virtual Observatory services, INAF-OATs or CANFAR located, the authorization policies are based on the user group memberships. On the other hand, in the EFC the authorization policies are related to the Virtual Organization membership. In our science platform, Virtual Observatory users must be mapped in a single specific Virtual Organization and the granularity of local data access policies is driven by the local groups memberships stored in the local Access Control Services. The access to shared cloud resources is driven by the specific Virtual Organization membership.

\paragraph{Proxy authentication in the Access Control}
The Access Control Service supports the proxy authentication mechanism by design, so EGI users having a X.509 certificate and belonging to a Virtual Organization can request a VOMS proxy and use it to access both OATs-INAF VObs gateway A\&A community specific services and EGI services. 

\paragraph{Access Control and Keystone-VOMS interoperability}
A user belonging to a Virtual Organization and registered both in the OATs-INAF cloud site and in the OATs-INAF Access Control, can access seamlessly cloud resources and VObs A\&A community specific services located at INAF-OATs exploiting the proxy authentication: a user, given his X.509 certificate and his Virtual Organization membership, can ask a VOMS proxy certificate and, using it, perform the authentication.

The ldap server used to store the Access Control Service users and the OpenStack cloud users in the OATs-INAF infrastructure is the same. This way, a user can access the IVOA services and the cloud resources also exploiting the same login/password authentication.

\paragraph{Authorization interoperability}

Data access authorization interoperability exploits proxy certificates, the certificate delegation concept and specific access control mechanisms.
The authorization information is stored in the VOSpace front-end as meta-data. Data can be public or private and, in case they are private, a set of groups of users have the read and/or write permissions. The group name is assumed to contain the information about which Access Control Service stores the information about a group (i.e. creator and group members). This is obtained using a group IVOA compliant identifier with the form: ivo://$<$ivoa-authority-identifier$>$/gms?$<$group-name$>$, containing the information about the authority (i.e. substantially the site) hosting the group.  

If a PI (Project Investigator) wants, for example, share with an INAF-OATs collaborator a private dataset, related to an observation, stored at CANFAR he has to perform the following steps:
\begin{itemize}
\item create a group in the gatway site INAF-OATs (e.g. ivo://oats.inaf.it/gms?experimentname)
\item ensure that his collaborator is registered as INAF-OATs gatway site user
\item add the collaborator as a new group member
\item give to the newly created INAF-OATs group the access rights (read and write permissions) to the CANFAR stored dataset 
\end{itemize}
At this point, the collaborator needing data access have to
\begin{itemize}
\item have a X.509 certificate
\item be member of a Virtual Organization
\item create a VOMS certificate proxy
\item access the INAF-OATs cloud gateway using the proxy certificate
\item instantiate a cloud virtual machine using the ad hoc image providing A\&A community VObs services clients
\item access the virtual machine with the same user-password credentials registered in the Access Control Service
\item copy his proxy in the virtual machine and delegate his credentials (proxy) to the CANFAR Credential Delegation Service
\item request the data to the CANFAR VOSpace Service
\end{itemize}
At this point, the CANFAR VOSpace Service knows the groups having access rights to that data, from the group names it knows which authority hosts that groups and so, getting the user's credentials from the CANFAR Delegation Service, it uses them to query the right Access Control Services to know the members of that groups. If the user requesting the data belongs to a group having access, the VOSpace Service uses the user's delegated credentials to retrieve and return the data to the querying user.

The interoperability between the differently located (Canada and Europe) storage services is performed through a test plan carried on jointly by CANFAR and OATs, specifically for what concerns the authentication, authorization and data movement aspects. Conditions and steps described in the previous paragraph have been verified.

\section{Conclusions and future work}
\label{conclusions}

The new generation of scientific instruments is starting to produce an unprecedented amount of data. At the same time, 
the recent  astrophysics and astroparticle experiments, although aiming at different scientific goals, 
have the common characteristic to be so complex to require the cooperation of large international 
collaborations collecting the expertise of Astronomers from different countries. 
Handling and exploring the new big data volumes in world-wide collaborations with the purpose of making real scientific discoveries, requires scientific research to be preceded by a considerable technical activity based on the adoption of new approaches. These shall overcome the traditional data processing methods and require the exploitation of new computing and storage facilities and science platforms for the access, visualization and analysis of the data. These challenges can be faced through the use of appropriate science platforms. 
With our work we have demonstrated the feasibility of building a science platform, by merging a cloud approach to computing and storage resource consumption with an integrated VObs implementation based on IVOA standards. In such a way, a FAIR (Findable, Accessible, Interoperable, Reusable) data access and management solution can be achieved. This science platform is deployed following an hybrid cloud model in which the CANFAR e-infrastructure and the EFC are interoperable, thanks to a common authentication model based on X.509 certificates and to the implementation of IVOA standards.
All the work done to deploy the Virtual Observatory compliant services is detailed, and on-line documentation is publicly available to disseminate our experience\footnote{https://github.com/oats-cadc}.

For the future, many plans are being made, involving developers, platform administrators and final users. Among technical aspects, the need is prominent to evaluate the feasibility of adopting new standards for authentication, authorization and credential delegation, like Identity Federation, OAuth, OpenId Connect. A thorough discussion on this topic has started within the IVOA environment. We plan to implement a container-based deployment procedure to simplify the adoption of our approach and to engage new institutions. From the final users perspective work is on-going to collect requirements for specific use cases and the provision of ad-hoc user interfaces is planned. 

Our work has shown that several cloud infrastructures, world-wide distributed, can be federated to create a global discipline-specific cloud, useful for each of the future big data projects and for interoperating among all of them. 
The ultimate goal of this activity is defining and implementing a technology to build a shared world-wide VO-compliant platform, based upon a number of facilities allowing (i) to access data and (ii) to perform processing on them. 
In such a way astrophysics and astroparticle physics would migrate from a community of managers/curators/distributors of data collections to a community capable of managing an integrated system of services for data-intensive astronomy. 

\section{Acknowledgements}
This work has been partially supported by the European Commission through the projects EGI-Engage (grant no. 654142) and ASTERICS (grant no. 653477), in the framework of the Horizon2020 research and innovation programme. The authors are indebted to Tiziana Ferrari (EGI) and David Schade (CADC) for many useful discussions.


\bibliography{egi_canfar_results}

\end{document}